\documentclass[dblblindworkshop]{article}

 \usepackage{comment}
 \usepackage{graphicx}
 \usepackage{amsmath}

  \usepackage[preprint]{neurips_2026}

\usepackage[utf8]{inputenc} 
\usepackage[T1]{fontenc}    
\usepackage{hyperref}       
\usepackage{url}            
\usepackage{booktabs}       
\usepackage{amsfonts}       
\usepackage{nicefrac}       
\usepackage{microtype}      
\usepackage{xcolor}         
\usepackage{tcolorbox}
\usepackage{float}
\usepackage{subcaption}
\title{Multi-Agent Orchestration with the Common-Sense Reasoning Capabilities of LLMs for Autonomous Driving}

\author{%
  Mehdi Azarafza\\
  Department of Computer Science \\
  Hamm-Lippstadt University \\ of Applied Sciences, Germany \\
  \texttt{mehdi.azarafza@hshl.de} \\
  \and
  Faezeh Pasandideh\\
  Department of Computer Science \\
  Hamm-Lippstadt University \\ of Applied Sciences, Germany \\
  \texttt{Faezeh.Pasandideh@hshl.de} \\
   \and
  Ali Ehteshami Bejnordi\\
  Department of Computer Science \\
  Hamm-Lippstadt University \\ of Applied Sciences, Germany \\
  \texttt{ali.ehteshami-bejnordi@hshl.de} \\
     \and
  Stefan Henkler\\
  Department of Computer Science \\
  Hamm-Lippstadt University \\ of Applied Sciences, Germany \\
  \texttt{stefan.henkler@hshl.de} \\
     \and
  Achim Rettberg\\
  Department of Computer Science \\
  Hamm-Lippstadt University \\ of Applied Sciences, Germany \\
  \texttt{achim.rettberg@hshl.de} \\
}

\begin{document}

\maketitle

\begin{abstract}
Autonomous vehicles require robust perception and decision-making capabilities to operate in diverse and unseen scenarios. While reinforcement learning and rule-based methods can provide effective control and safety mechanisms, their performance may degrade in situations requiring contextual reasoning. Large Language Models (LLMs) have demonstrated strong capabilities in understanding multimodal information and generating contextual reasoning, however, their use for direct vehicle control can introduce latency and hallucination risks. To address these limitations, a hybrid framework is proposed in which LLMs are used offline. The system uses an orchestrator to coordinate PPO-trained reinforcement learning and rule-based planning with PID control, while LLM common-sense reasoning is applied throughout the framework. LLM reasoning is further employed iteratively to refine the RL reward function for dynamic driving environments. The proposed framework is evaluated in highly randomized CARLA scenarios under diverse environmental and traffic conditions. The results demonstrate the potential of integrating LLM-based reasoning with conventional autonomous driving methods while retaining structured control and safety mechanisms.
\end{abstract}

\section{Introduction}
In recent years, autonomous vehicles have advanced significantly. However, the complexity of real-world environments remains a major challenge for safe and reliable driving \cite{verdi}.
Despite recent progress, Reinforcement Learning (RL) remains limited by its dependence on reward-driven optimization, which provides no formal guaranties of safe behavior and often restricts generalization beyond trained environments \cite{mastermind}. Rule-based systems are constrained by predefined logic and struggle in dynamic conditions. In contrast, Large Language Models (LLMs) demonstrate better flexibility and stronger cross-domain generalization by integrating diverse contextual information \cite{mastermind}
, although their use in real-time control remains challenging.\\
Furthermore, human driving requires not only vehicle control, but also visual perception and deductive reasoning, the ability to infer hidden or future states, perform counterfactual reasoning, and distinguish between visually similar but semantically different scenarios \cite{kothawade2021autodiscern}. These capabilities highlight the need for autonomous driving systems that can combine low-level vehicle control with contextual reasoning and decision-making under diverse and uncertain driving conditions. Table \ref{tab:method_comparison} presents a comparison of these methods. Each model is associated with specific limitations. Rule-based methods rely on predefined rules and scenarios, which can limit their adaptability to highly dynamic environments. RL has demonstrated stronger performance in such environments due to its capacity for learning and adaptation from feedback; however, RL methods often face challenges in generalizability \cite{mastermind}. LLMs have demonstrated strong capabilities in dynamic settings and greater generalizability, but their response latency may limit their effectiveness in real-time applications \cite{anonymous2024hybrid}.\\
To address these challenges, a hybrid approach is proposed that combines rule-based methods and reinforcement learning while leveraging the reasoning capability of LLMs through an orchestrator. Instead of using the LLM as a direct decision-maker in real time (due to latency and potential hallucination issues) we employ it as an advisory component and as a reward mechanism for the RL model. In this work, real-time operation is defined as responding within a fixed time deadline.
This design enables improved reasoning support while maintaining real-time performance and system reliability. 

\begin{table}
  \caption{Comparison of LLM, RL, rule-based, and hybrid decision-making methods in autonomous driving}
  \label{tab:method_comparison}
  \centering
  \begin{tabular}{lcccc}
    \toprule
    \textbf{Criteria} &
    \textbf{LLM$^{\mathrm{a}}$} &
    \textbf{RL$^{\mathrm{b}}$} &
    \textbf{Rule-based} &
    \textbf{Hybrid} \\
    \midrule
    Decision-making & $\checkmark$ & $\checkmark$ & $\checkmark$ & $\checkmark$ \\
    Flexibility & $\checkmark$ & $\times$ & $\times$ & $\checkmark$ \\
    Generalization & $\checkmark$ & $\times$ & $\times$ & $\checkmark$ \\
    Real-time Response & $\times$ & $\checkmark$ & $\checkmark$ & $\checkmark$ \\
    Common-Sense Reasoning & $\checkmark$ & $\times$ & $\times$ & $\checkmark$ \\
    \bottomrule
  \end{tabular}

  \vspace{2mm}
  \footnotesize
  $^{\mathrm{a}}$Large Language Model, 
  $^{\mathrm{b}}$Reinforcement Learning.
\end{table}

\begin{figure}
  \centering
  \includegraphics[width=1\textwidth]{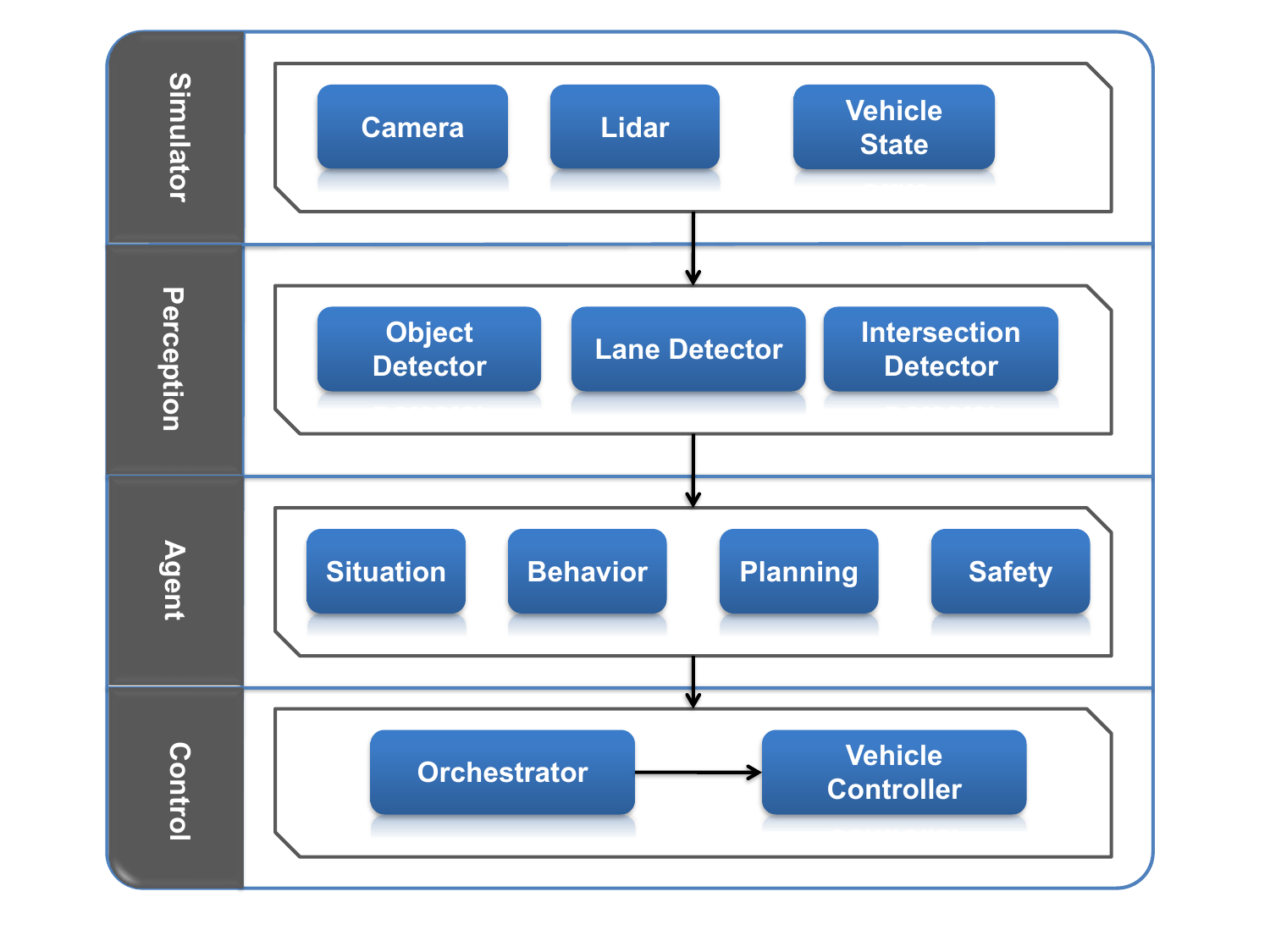}
  \caption{High-level system architecture}
  \label{fig:architecture}
\end{figure}

Figure \ref{fig:architecture} illustrates the high-level system architecture, showing the organization of its components.
The hybrid architecture is structured in multiple layers, beginning with the perception layer implemented in CARLA \cite{carla}, where camera and LiDAR sensors are employed for environmental sensing. The knowledge base layer incorporates the LLM as both a reasoning module and a reward model.
This is followed by a multi-agent layer, and finally the driving mode layer, which generates the control actions.

\section{Related Work}

Integrating LLMs into autonomous driving RL policies as direct decision-makers risks hallucination and unsafe execution. To mitigate this, HCRMP \cite{chen2025hcrmp} introduces LLM-Hinted RL, restricting the LLM to generating semantic state hints while bridging low-frequency reasoning and high-frequency control via memory caching and contextual anchoring. However, HCRMP struggles during dynamic traffic density shifts and lacks real-world validation beyond idealized CARLA simulations.

Alternatively, Li et al.  \cite{li2025llm} adopt a hierarchical architecture where the LLM generates long-term goals and meta-actions, while a low-level RL agent executes continuous control (steering and acceleration). They align policy updates via a Goal Gradient-based Transfer mechanism and refine rewards through human preferences. Nevertheless, LLM-generated goals remain vulnerable to unsafe hallucinations in edge cases, and the framework’s evaluation is constrained by a small human evaluation pool and limited scenario diversity.

Diverging from LLM-centric designs, Choi and Kim \cite{choi2025predictive} reformulate the intrinsic RL risk signal without LLMs. They introduce Safety Potential, a dense reward-shaping metric based on predicted path overlap and temporal weighting. While effective in reducing collisions in CARLA, this approach is limited to longitudinal control (relying on a rule-based Pure Pursuit controller for steering), depends on ground-truth simulator states, and underpredicts risk during complex diagonal vehicle interactions. Closest to hybrid safety integration, LSADQN \cite{ren2026llm} screens experience data via physical criteria (e.g., TTC), invoking an LLM (GPT-4) only for ambiguous samples to enforce margin-based contrastive regularization in value learning. Although it achieves < 1\% collision rates in highway-env, invoking the LLM online still introduces computational overhead and latency bottlenecks. Furthermore, its heuristic safety thresholds lack formal guarantees, urban applicability, and multi-agent validation.

Existing literature presents a fundamental trade-off: online LLM integration introduces latency and hallucination risks, whereas pure RL models suffer from single-dimensional control, over-reliance on ground-truth states.

We address these limitations with an Orchestrator-based multi-agent framework that completely decouples the LLM from real-time execution. The LLM operates strictly offline for reward refinement (with human validation) and static rule extraction. During runtime, risk is dynamically evaluated by a multi-factored Safety Agent  and Situation Agent. Crucially, the Safety Agent possesses a deterministic ASIL-D-compliant veto, triggering emergency overrides when risk thresholds are breached. At the control level, two independent PPO networks learn joint longitudinal and lateral control from raw camera (YOLOv11) and LiDAR inputs, ensuring end-to-end traceability and industry-aligned safety.

\section{Methodology}
\begin{figure}[H]
    \centering
    \includegraphics[width=\textwidth]{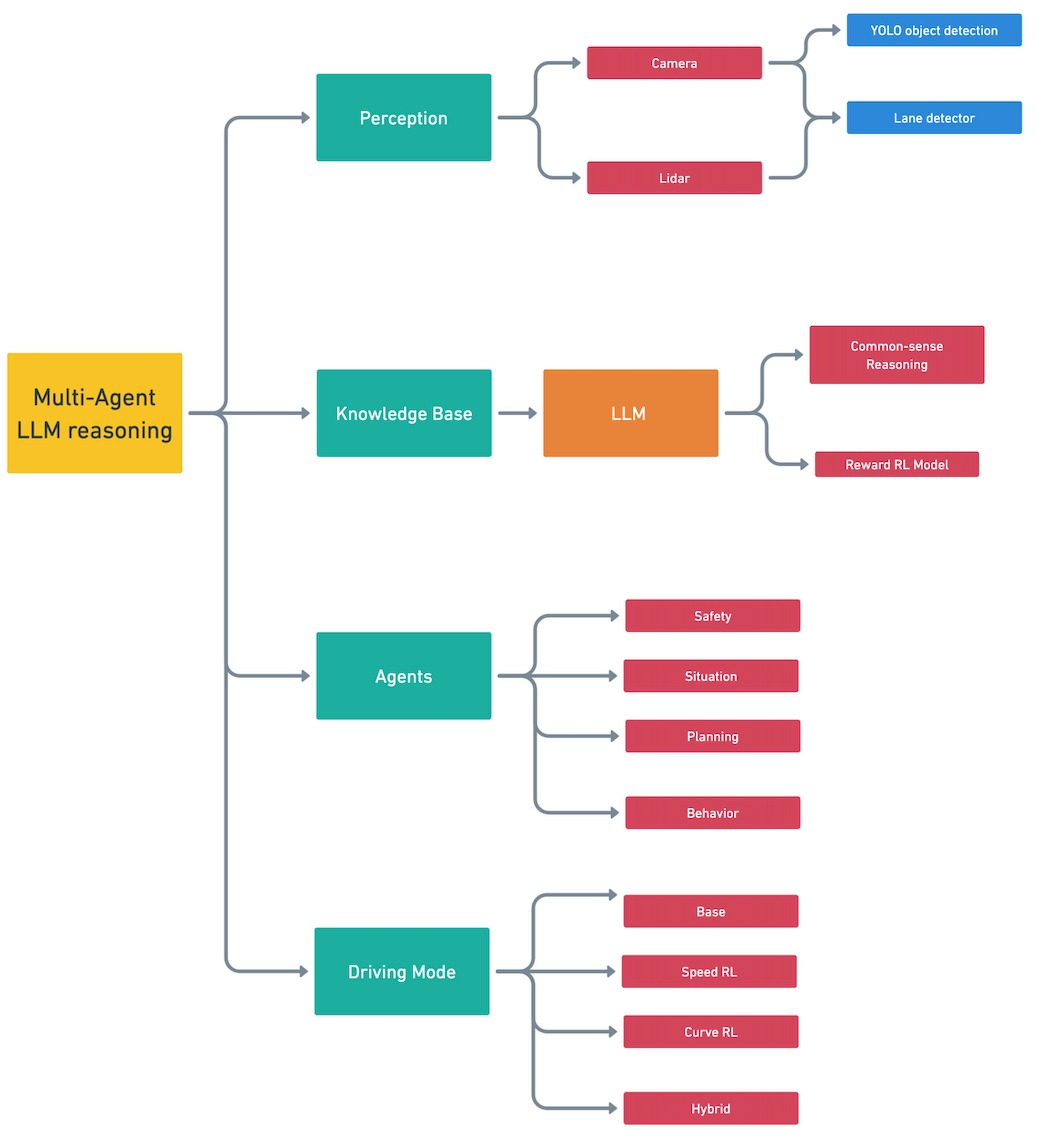}
    \caption{Hierarchical system architecture of the proposed multi-agent framework}
    \label{fig:system_architecture}
\end{figure}

The proposed framework operates in two domains: an offline development phase and a runtime execution phase. In the offline phase, LLM-based reasoning guides reward function refinement, while in runtime a multi-agent architecture with orchestrator controls the vehicle. Figure~\ref{fig:system_architecture} provides a detailed overview of the
hierarchical system architecture of the proposed multi-agent framework.

At runtime, camera and LiDAR data are processed in the perception layer to generate a unified perception state comprising object detections, distances, lane geometry, and environmental context. This state serves as shared input to four specialized agents operating in parallel: a Safety Agent (ASIL-D) for collision risk and rule compliance assessment, a Situation Agent (ASIL-C) for contextual analysis, a Planning Agent (ASIL-B) for target speed selection, and a Behavior Agent (ASIL-A) for comfort optimization. Each agent produces an action recommendation with associated confidence and reasoning trace. A central decision arbitrator performs emergency checks and applies ASIL-priority–based fusion, ensuring that higher safety levels dominate in case of conflict. The final decision specifies steering and speed commands with full traceability. Control execution is handled by rule-based planners paired with PID controllers or PPO-trained RL policies, with safety constraints overriding learned behavior when necessary. During development, the LLM (GPT 5.2)\cite{gpt5} summarizes and analyzes logged episodes to identify failure patterns and recommend reward adjustments. Human validation precedes retraining, forming an iterative improvement cycle. A knowledge base of approximately 100 common-sense rules from Claude \cite{claude} supports both runtime safety enforcement and reasoning. In the following subsections. We present a detailed description of each layer and component of the proposed framework.

\subsection{Perception}
The perception layer integrates three complementary modules to generate a comprehensive representation of the environment. A YOLOv11 model \cite{yolo}, trained in Roboflow for CARLA-relevant object detection, processes camera images to detect and classify traffic participants (including vehicles, pedestrians, cyclists and traffic lights) along with associated confidence scores. The lane detection module employs edge detection and polynomial fitting to extract lane boundaries, calculating center offsets for lateral positioning and curvature estimation. The intersection detector evaluates road geometry by estimating lane widths and identifying horizontal markings such as crosswalks. Outputs from all three modules are fused into a unified perception state that captures both dynamic objects and static road structures, forming the foundational input for subsequent reasoning layers.
\subsection{Agents}
The multi-agent reasoning layer employs four specialized agents operating in parallel, each assigned a distinct safety integrity level. Claude Sonnet\cite{claude} were employed as advisory models for safety and regulatory alignment. Each agent produces a recommendation containing target speed, steering offset, stop command, driving mode suggestion, and confidence score.
In addition, a set of structured assumptions was adopted to operationalize requirements within the proposed multi-agent framework. These assumptions were established based on LLM-generated recommendations and human review, and reflect the design choices made in our implementation. \\

\textbf{The Safety Agent}: rated at ASIL-D as the highest criticality level,
serves as the primary guardian of vehicle safety. It estimates time-to-collision
(TTC) from consecutive front-distance measurements and relative velocity. When
an approaching object is detected, TTC is computed from the closing distance and
relative velocity, an ego-velocity-based
estimate is used. Collision risk is then classified using multiple TTC thresholds:
emergency ($<1.5$\,s), warning ($<3.0$\,s), caution ($<5.0$\,s), and low risk
otherwise. The resulting risk level is used by the orchestrator to prioritize
safety-critical actions.

This risk is further adjusted based on absolute distance thresholds, taking the maximum of TTC-based risk and distance-based risk where $R_{collision} = \max(R_{collision}, 0.9)$ when front distance falls below emergency brake distance.

The agent integrates a common-sense reasoning engine containing over thousand driving rules. From these LLM-generated rules, based on CARLA capabilities, only a subset can be applied, therefore, approximately 100 rules were selected for implementation. These rules cover traffic signals, pedestrians, weather conditions, and defensive driving principles. The agent evaluates the current perception state against these rules and selects the most critical triggered rule for decision-making. It performs holistic risk assessment combining multiple factors:
\begin{equation}
\begin{split}
R_{\text{holistic}} = \min \big( & 0.5 \cdot R_{\text{collision}} 
+ 0.1 \cdot N_{\text{faults}} \\
& + 0.2 \cdot L_{\text{perception}} 
+ R_{\text{traffic}} 
+ R_{\text{intersection}}, 1.0 \big)
\end{split}
\end{equation}\\
where $N_{faults}$ represents the count of sensor faults, $L_{perception}$ denotes perception limitation score, $R_{traffic} = 0.15$ when traffic light is red, and $R_{intersection} = 0.1$ when at an intersection. Based on this assessment, the agent determines appropriate safe states: normal operation when $R_{holistic} \leq 0.4$, degraded operation when $0.4 < R_{holistic} \leq 0.7$, minimal risk condition when $R_{holistic} > 0.7$, and safe stop when $R_{collision} > 0.9$.\\\\
\textbf{The Situation Agent}: rated at ASIL-C, provides environmental context
assessment and evaluates Operational Design Domain (ODD) compliance based on
speed constraints, weather, lighting, road type, sensor health, and traffic
density, to the extent these factors can be implemented within CARLA. Current
conditions are classified according to the number of detected ODD violations:
normal, edge case , degraded, and out of ODD.
The situation risk is then determined from collision probability, the number of
ODD violations, scenario category, and traffic conditions.
maximum risk of 1.0.\\
\textbf{The Planning Agent}: rated at ASIL-B, handles strategic decision-making including action planning and turn decisions. It maintains a road model tracking current lane position and intersection status, plans actions based on current scenario, and employs rule-based turn decisions using distance-weighted scoring with exploration factors and turn history penalties. The agent supports optional RL-based learned turn decisions when learning mode is explicitly enabled, though this capability is disabled by default in the current implementation to ensure deterministic behavior for safety validation.\\
\textbf{The Behavior Agent}: rated at ASIL-A, optimizes driving quality and comfort. It assesses driving quality through smoothness metrics based on steering variance calculated from a sliding window of recent center offsets.\\ 
\subsection{Orchestrator} 
The orchestrator manages multi-agent recommendations, with the decision arbiter resolving conflicts using ASIL-weighted priority fusion. For each agent $i$, the effective weight is calculated as:

\begin{equation}
w_i = w_{base,i} \cdot m_{ASIL,i} \cdot c_i \cdot b_{conflict,i}
\end{equation}

where $w_{base}$ denotes the base agent weights (Safety: 2.5, Situation: 1.5, Planning: 1.2, Behavior: 0.8), and $m_{ASIL}$ denotes the corresponding ASIL multipliers (D: 2.0, C: 1.5, B: 1.2, A: 1.0). The confidence score is denoted by $c_i$, while $b_{conflict}$ is set to 1.3 for agents that win conflict resolution and to 1.0 otherwise. These weights and safety priorities were selected as design assumptions for urban driving scenarios, with CARLA Town02 being used as the primary environment. The weighting was selected to reflect the higher density of interaction and the safety requirements associated with urban driving. Extension to highway or Autobahn scenarios may require further adjustment of these priorities.
When agents conflict on stop decisions, any ASIL-D or ASIL-C agent recommending stop with confidence above threshold has veto power that forces immediate stop regardless of other agent opinions.

The proposed multi-agent orchestration framework comprises a priority-weighted arbiter and four specialized sub-agents (see Appendix, Figure~\ref{fig:orchestrator}).
Table~\ref{tab:control_strategies} summarizes the control strategies evaluated in this work, covering baseline PID control and configurations incorporating reinforcement learning for curve and speed control. The hybrid strategy combines PID and RL components to provide control across both lateral and longitudinal behaviors.
\begin{table}
  \caption{Comparison of control strategies used in autonomous driving.}
  \label{tab:control_strategies}
  \centering
  \begin{tabular}{lcccc}
    \toprule
    \textbf{Strategy} & \textbf{Steer} & \textbf{Curve} & \textbf{Speed} & \textbf{Reasoning} \\
    \midrule
    Base   & PID    & PID & N/A & Common sense \\
    Curve  & PID    & RL  & N/A & Common sense \\
    Speed  & PID    & N/A & RL  & Common sense \\
    Hybrid & PID+RL & RL  & RL  & Common sense \\
    \bottomrule
  \end{tabular}
\end{table}

\subsection{Reinforcement learning}

In this work, reinforcement learning was applied as part of the control module in three modes. Separate RL models were used for speed control and curve handling, while a combination of both was utilized in the hybrid mode. Two independent PPO-based actor-critic networks \cite{ppo} handle control, respectively. Both employ the same hidden architecture and identical training hyperparameters. The orchestrator invokes them independently at each control tick, and the safety agent retains override authority over both outputs.

The steering controller outputs $\delta \in [-1, 1]$, the same value of steering control the vehicle steering from CARLA \cite{carla_steering}, for lane keeping and curve following. The reward signal penalises lane deviation and steering discontinuities. The key state variables used by the RL controllers are provided in ~\ref{tab:rl_state_vectors}.

\begin{table}[H]
  \caption{Key state features used by the RL speed and curve controllers.}
  \label{tab:rl_state_vectors}
  \centering
  \small
  \begin{tabular}{lll}
    \toprule
    \textbf{Controller} & \textbf{State Variable} & \textbf{Value/Scale} \\
    \midrule
    RL Speed & Speed & $v/50$ \\
    & Target Speed & $v_{\mathrm{target}}/50$ \\
    & Front Distance & $d/50$ \\
    & Front-Left / Front-Right Distance & $d/30$ \\
    \midrule
    RL Curve & Center Offset & $o/2$ \\
    & Curvature & $10\kappa$ \\
    & Speed & $v/50$ \\
    & Front Distance & $d/50$ \\
    \bottomrule
  \end{tabular}
\end{table}

\section{LLM Integration}
This section describes the integration of LLMs into two components of the proposed framework. First, an LLM is employed as a reward refinement mechanism for the RL-based curve controller. The initial RL model is trained and evaluated without LLM guidance, after which its outputs are analyzed by GPT-5.2 to identify potential improvements. Second, Claude is incorporated into the common-sense reasoning component. The following subsections describe these two applications in detail.
\subsection{LLM as a Reward Refinement Mechanism for RL}
An initial RL model (V0) was trained without LLM guidance using data collected from CARLA over 1,000 episodes. The episodes were conducted in highly dynamic environments without predefined scenarios, allowing vehicles and pedestrians to appear at varying locations and under different traffic conditions. This setup was intended to expose the controller to various situations, including configurations that may not have been adequately represented during training.

The reward function was iteratively refined using GPT-5.2 feedback based on summarized curve-steering episodes. The initial refinement improved some steering metrics but reduced overall progress, motivating further adjustment of the reward scale and penalty terms. The complete iterative refinement process is provided in Appendix~\ref{sec:llm_rl_refinement}.\\
\subsection{LLM as a Common-Sense Reasoner}

A common-sense reasoning module was implemented using Claude as the LLM to
generate driving rules for diverse environmental, traffic, and road-user
conditions. The generated rules covered scenarios including traffic-light
states, weather, night-time driving, vehicle following distance, pedestrians,
cyclists, and intersections. The rules were subsequently mapped to executable
conditions and actions within CARLA. Rules involving actions that were not
supported or observable in the simulator, such as horn and flashing-light
behavior, were excluded, while rules with corresponding executable conditions
and actions were retained. Approximately 100 executable rules were generated and integrated into CARLA. 

The resulting rule set provides context-dependent driving behavior beyond
basic traffic-light handling. For example, red-light conditions can trigger
stopping or gradual speed reduction based on distance, while adverse weather
can increase caution and reduce speed. Similarly, close vehicle following can
trigger an emergency stop, whereas following in rain can require increased
spacing. Pedestrian-related rules include stopping at crosswalks, yielding at
intersections, and reducing speed for pedestrians under reduced-visibility
conditions. Additional rules address cyclists, night-time driving, and
weather-dependent intersection behavior.

The generated rules were integrated into the \textit{Safety Agent}, which was
assigned the highest priority within the orchestration framework. When a
corresponding condition is detected, the Orchestrator can invoke the
common-sense reasoner and forward the selected action to the CARLA control
system. This integration allows contextual rules generated by the LLM to be
translated into executable safety-oriented behavior while retaining the
Safety Agent's priority over other control decisions.

\begin{figure}
    \centering
    \begin{subfigure}{0.48\textwidth}
        \centering
        \includegraphics[width=\textwidth]{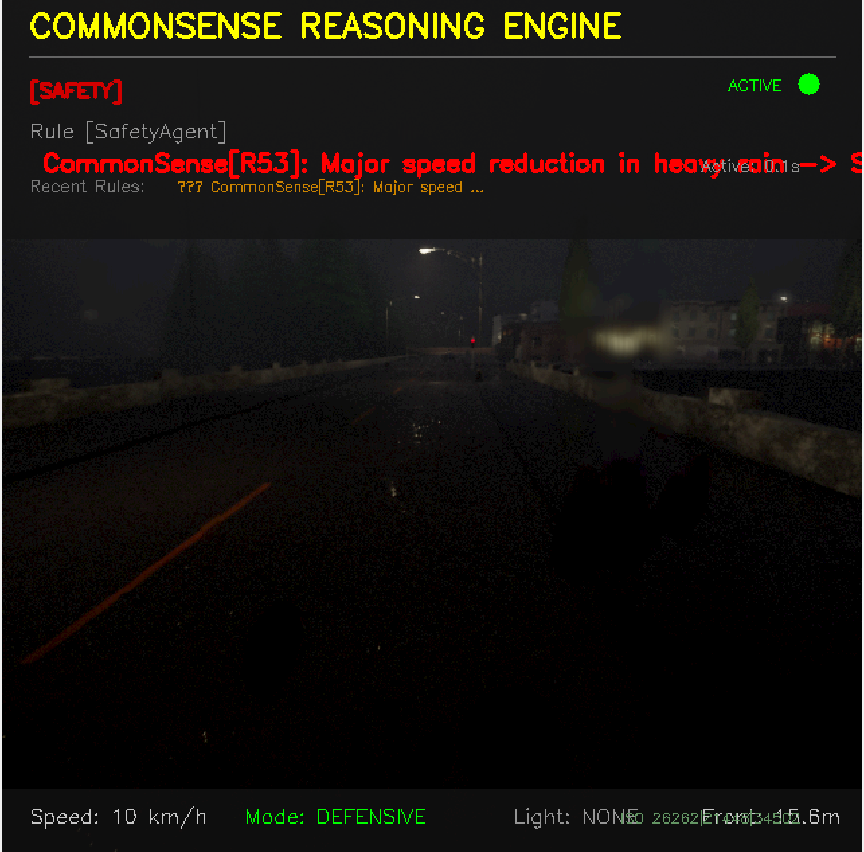}
        \caption{}
        \label{fig:heavy_a}
    \end{subfigure}
    \hfill
    \begin{subfigure}{0.48\textwidth}
        \centering
        \includegraphics[width=\textwidth]{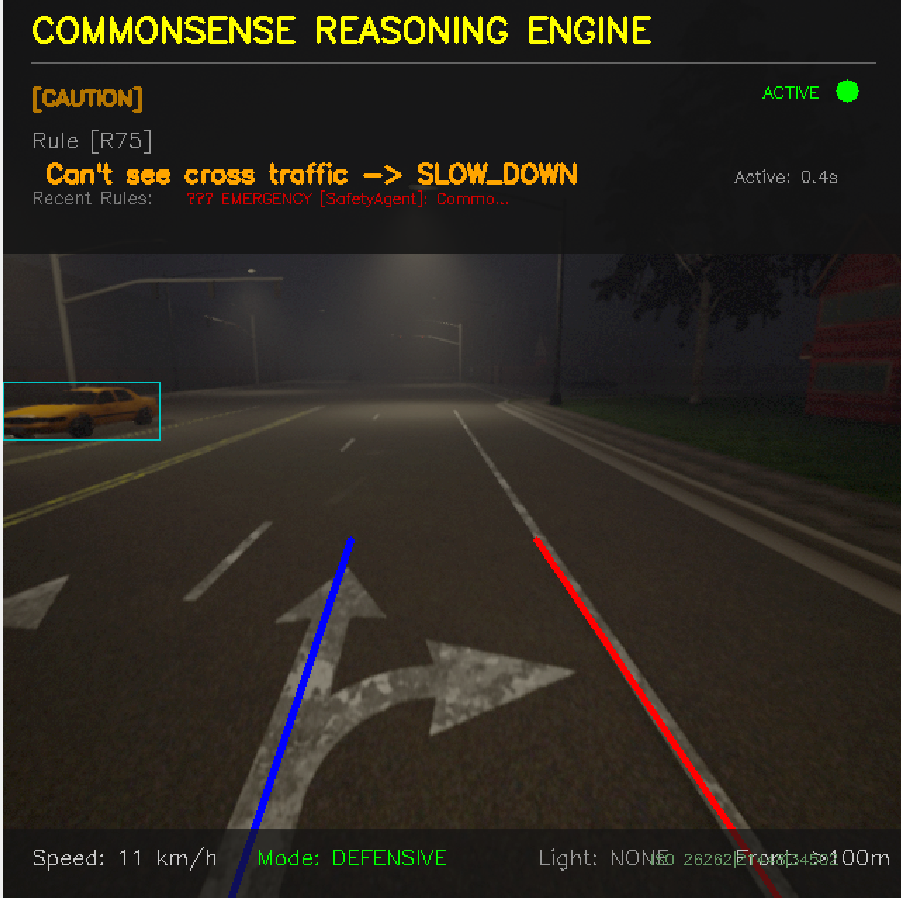}
        \caption{}
        \label{fig:cant_see}
    \end{subfigure}

  \caption{Examples of common-sense reasoning in CARLA. (a) Rule~53: speed reduction under heavy rain. (b) Rule~75: speed reduction when cross traffic cannot be observed}
    \label{fig:common_sense}
\end{figure}
Figure~\ref{fig:common_sense} illustrates two examples of common-sense reasoning under driving conditions in CARLA. Figure~\ref{fig:common_sense}(a) shows Rule~53, which requires a substantial speed reduction under heavy rain and results in the Safety Agent reducing the ego vehicle's speed. Figure~\ref{fig:common_sense}(b) shows Rule~75, where limited visibility of cross traffic triggers a speed reduction for the ego vehicle. Additional examples are provided in Appendix~\ref{sec:common_sense_examples}.
Figure~\ref{fig:appendix_common_sense}(a) shows a short following distance
that triggers a slow-down command (Rule~156), with the reasoning linked to an
approaching storm (Rule~93). Figure~\ref{fig:appendix_common_sense}(b) shows
a stop signal at a red traffic light in heavy rain.
\section{Evaluation}
As discussed in Section 4.1, four different RL versions were developed, with iterative feedback from the LLM used to refine the models over time. V0 was trained without LLM feedback, while V3 represents the model obtained after three rounds of LLM-based feedback and refinement. The complete iterative refinement process is provided in Appendix~\ref{sec:llm_rl_refinement}. 

Figure~\ref{fig:my_curve} compares the learning dynamics of V0–V3 using trailing 50-episode statistics. Although all variants exhibit non-monotonic behavior, V3 provides the strongest end-of-training balance between tracking accuracy, steering stability, and curve progress. At episode 1,000, V3 achieves the lowest median lane error (0.0514) and the lowest curve-oscillation rate among the non-degenerate variants (45.89\%), while maintaining the highest median curve exposure (193.5 steps). V2 attains comparable exposure (187.5 steps) but produces higher lane error (0.0849), greater steering-change magnitude (0.3708), and a higher oscillation rate (47.99\%). V1 remains comparatively stable but reaches only 101.5 curve steps and exhibits the highest terminal oscillation rate (49.69\%). V0 shows a sharp late reduction in steering changes and oscillations; however, this coincides with its curve exposure collapsing to 21.5 steps, indicating reduced driving opportunity rather than genuine stabilization. Overall, the results demonstrate that V3 achieves the most favorable balance between steering quality and sustained curve progress across the evaluated variants.

\begin{figure}[h]
    \centering
    \includegraphics[width=0.8\textwidth]{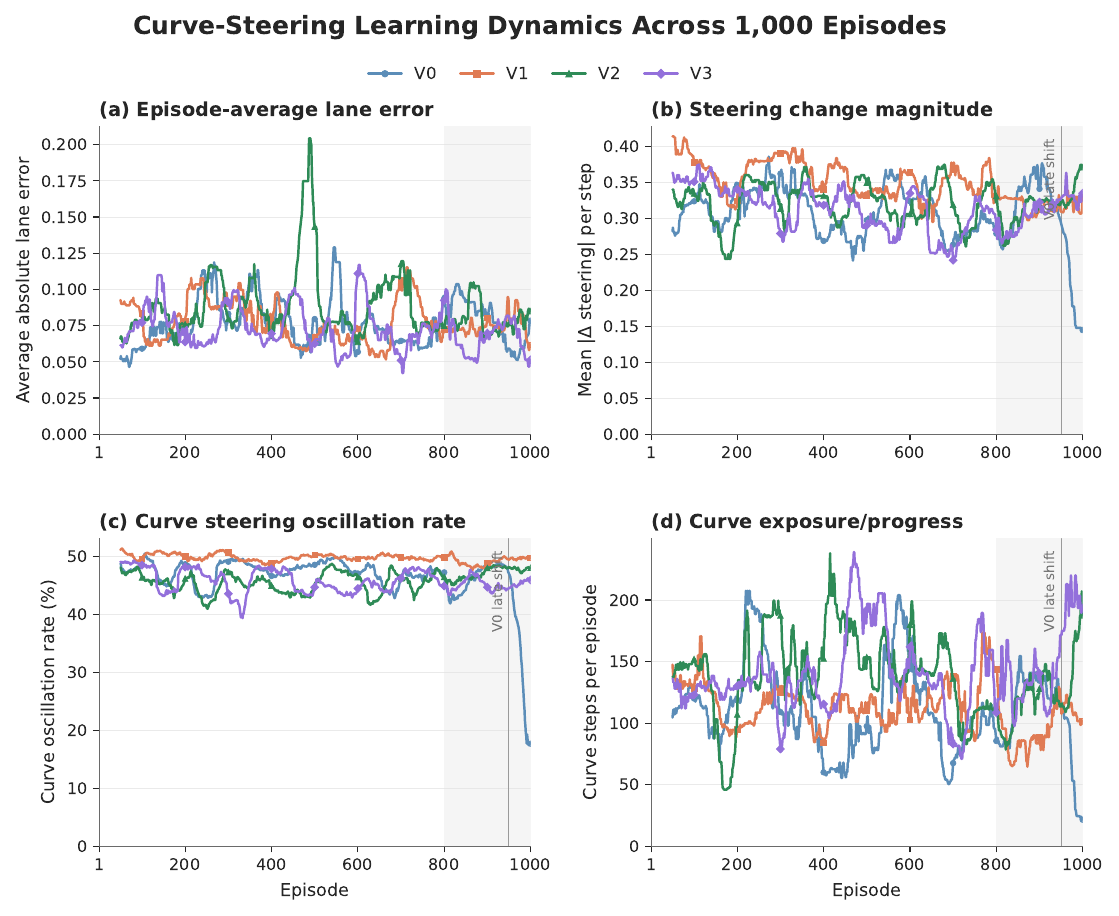}
    \caption{Training behavior of the four curve-steering reward variants}
    \label{fig:my_curve} 
\end{figure}

\begin{figure}[h]
    \centering
    \includegraphics[width=0.8\textwidth]{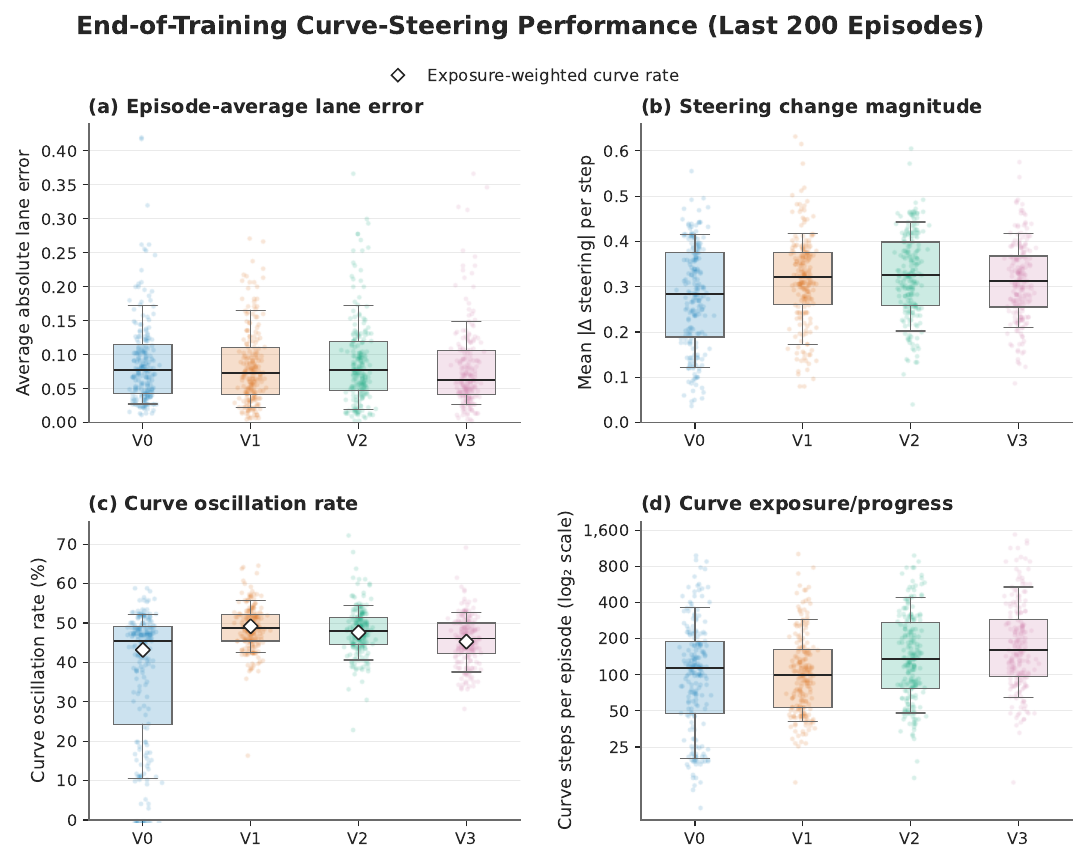}
    \caption{End-of-training assessment of tracking accuracy, steering stability, and curve progression across the four reward variants}
    \label{fig:my_curve2} 
\end{figure}

Figure~\ref{fig:my_curve2} summarizes the end-of-training behavior of V0–V3 over the final 200 episodes, where boxes represent the median and interquartile range, whiskers indicate the 10th–90th percentiles, individual points denote episode-level observations, and diamonds in panel (c) show exposure-weighted curve-oscillation rates. V3 achieves the lowest median lane error at 0.0627, corresponding to an improvement of approximately 15–20\% relative to V0, V1, and V2, whose medians are 0.0780, 0.0733, and 0.0770, respectively. In terms of steering-change magnitude, V3 obtains a median of 0.3119, improving upon V1 and V2 while remaining slightly higher than V0. A similar pattern is observed for curve oscillations: V3 records an exposure-weighted rate of 45.27\%, compared with 49.18\% for V1 and 47.61\% for V2, whereas V0 reaches a lower rate of 43.18\%. However, panel (d) shows that V0 operates under substantially lower curve exposure, with a median of 113 curve steps per episode, compared with 160 for V3. V3 also exceeds V1 and V2, which achieve medians of 99 and 135.5 curve steps, respectively. Consequently, the lower steering-change and oscillation values observed for V0 must be interpreted together with its reduced curve traversal. Overall, V3 provides the strongest combined performance, achieving the lowest lane error, improved steering stability relative to V1 and V2, and the greatest sustained curve exposure among all evaluated variants.

\section{Conclusion}

In this work, a hybrid autonomous driving framework is developed by combining a common-sense reasoner, PID  controllers, real-time object detection, and reinforcement learning. Four specialized agents and dedicated controllers were coordinated through an orchestrator, with decision authority retained by the orchestration framework rather than being directly delegated to the LLM. The framework was evaluated in CARLA under diverse and highly randomized scenarios, including different weather conditions, with the reasoning layer playing a particularly important role in decision-making under adverse conditions.

LLM-guided reward refinement was also employed iteratively to improve the RL controller. Across successive RL versions, improvements in the balance between tracking accuracy, steering stability, and curve progress were observed, demonstrating the potential of combining LLM reasoning with conventional autonomous driving approaches. Future work can investigate more advanced agentic AI architectures and explore how greater decision-making authority can be safely delegated to individual agents within constrained control settings.

\begin{ack}
The authors acknowledge the use of LLMs as a core methodological component of this work. In particular, LLMs were employed to explore common-sense reasoning capabilities for autonomous driving, to support alignment with relevant standards to the extent possible within CARLA, and as reward models within the proposed framework, with Claude and GPT used as part of the methodological contribution. The authors used generative AI tools strictly for grammar checking and copy-editing to improve the readability of the manuscript. The core methodology, structure, experimental design, analysis, and conclusions were developed entirely by the human authors.  Additionally, the system architecture diagram in this work was designed using Whimsical.



\end{ack}

{
\small
\bibliographystyle{plainnat}
\bibliography{references}
}

\appendix
\newpage

\section{Multi-Agent Orchestration}
\label{sec:appendix_orchestration}

\begin{figure}[H]
  \centering
  \includegraphics[width=\textwidth]{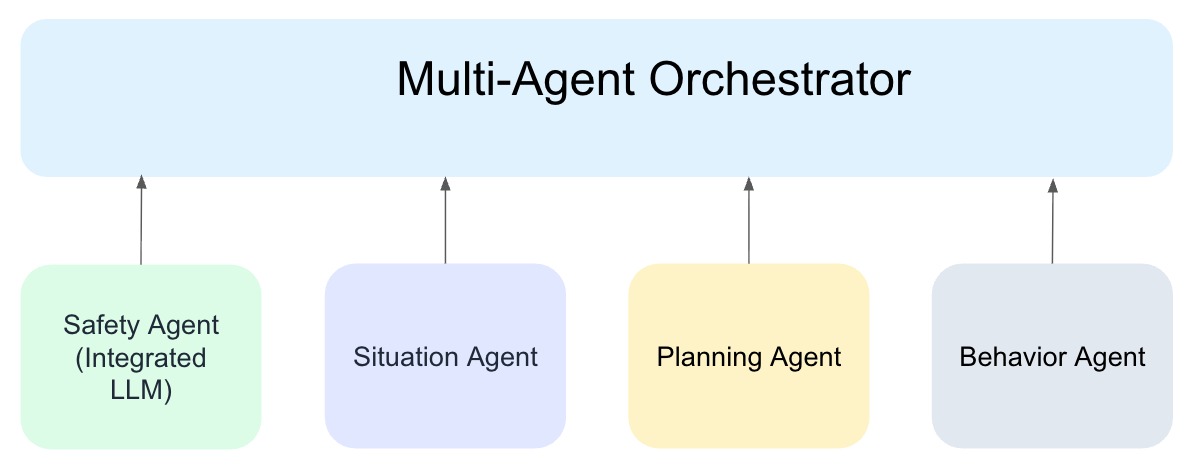}
  \caption{High-level overview of the proposed multi-agent orchestration with agents}
  \label{fig:orchestrator}
\end{figure}


\newpage
\section{LLM-Guided RL Reward Refinement}
\label{sec:llm_rl_refinement}

This section presents the iterative LLM-guided refinement of the RL reward function across successive model versions.

\begin{tcolorbox}[
  colback=white,
  colframe=black,
  arc=2mm,
  title={Part of LLM-guided reward refinement for curve steering},
  label={box:llm_rl_refinement}
]
\textbf{Prompt for V0:}\\

You are evaluating SUMMARIZED curve steering episodes for RL reward.
This is ONLY for CURVE STEERING RL (not speed control).

\begin{itemize}
    \item Focus on: steering quality, lane\_error, smoothness, and curve tracking.
    \item Ignore: speed metrics (handled by separate RL model).
\end{itemize}

\textbf{Key Metrics:}
\begin{itemize}
    \item \texttt{steering\_oscillations}: number of steering sign changes (higher = more oscillatory).
    \item \texttt{steering\_smoothness}: average $|\text{steering change}|$ per step (lower = smoother).
    \item \texttt{curve\_steering\_oscillations}: steering sign changes only during \texttt{is\_curve=True}.
    \item \texttt{avg\_lane\_error}: average absolute lane deviation.
    \item \texttt{outcome}: \texttt{collision}, \texttt{off\_road}, \texttt{good\_progress}, or \texttt{completed}.
\end{itemize}

\rule{\linewidth}{0.4pt}
\textbf{GPT-5.2 Evaluation:}

\textbf{Worst Problem:} \textit{curve\_tracking}

\textbf{Reward Adjustments:}
\begin{itemize}
    \item \textit{lane\_keeping} = 0.75
    \item \textit{smoothness} = 0.80
    \item \textit{curve\_tracking} = 0.85
    \item \textit{safety} = 1.00
    \item \textit{progress} = 0.25
\end{itemize}

\textbf{Priority Fixes:}
\begin{enumerate}
    \item Strongly penalize steering oscillations during \texttt{is\_curve}
    and add a penalty on steering jerk (\texttt{delta-steer}) to suppress
    rapid sign changes and dithering in bends.

    \item Increase lane-centering penalties with a non-linear term,
    applying stronger weighting during curves.

    \item Add an explicit curve-anticipation/heading-alignment shaping term
    to reduce late corrections that cause overshoot and run-off.
\end{enumerate}

\rule{\linewidth}{0.4pt}
\textbf{V1: Implementation of LLM Feedback}

The following reward modifications were implemented based on the V0
evaluation:

\begin{itemize}
    \item Added an oscillation penalty with stronger weighting during curves.
    \item Added a steering jerk penalty based on the second derivative.
    \item Changed the lane-deviation penalty from linear to quadratic.
    \item Added a curve-anticipation reward using a far waypoint at 12\,m.
    \item Applied the LLM-recommended weights:
    \textit{lane} = 0.75, \textit{smoothness} = 0.80,
    \textit{curve} = 0.85, \textit{safety} = 1.00,
    \textit{progress} = 0.25.
\end{itemize}
\end{tcolorbox}
\newpage

\begin{tcolorbox}[
  colback=white,
  colframe=black,
  arc=2mm,
  title={Part of V1 Evaluation and V2 Reward Refinement}
]

\textbf{GPT-5.2 Evaluation:}

\textbf{Root Cause:}
\textit{Reward scale/shape imbalance: penalty terms (especially jerk/oscillation and quadratic lane error) appear unbounded or too heavily weighted, dominating the return and pushing the policy toward behaviors that reduce penalty accumulation (low progress, early crash/off-road, or timid/oscillatory micro-corrections) rather than learning a stable steering law that matches curvature.}

\textbf{V2 Reward Adjustments:}
\begin{itemize}
    \item \textit{lane\_keeping} = 0.60
    \item \textit{smoothness} = 0.25
    \item \textit{curve\_tracking} = 0.90
    \item \textit{safety} = 1.00
    \item \textit{progress} = 0.80
\end{itemize}

\textbf{Priority Fixes:}
\begin{enumerate}
    \item Rescale and cap penalties
    
    \item Replace/augment oscillation counting with a bounded frequency/energy proxy
    
    \item Improve curve-tracking reward signal
\end{enumerate}

\textbf{summary:}
V1 slightly reduced lane deviation and oscillation counts but is overall worse because it removed all successes and significantly reduced route progress while producing extremely negative returns. The dominant issue is reward magnitude imbalance: new penalties swamp the learning signal and encourage avoidance/early failure rather than stable curve tracking. V2 should keep curve-specific shaping but heavily cap/normalize penalties and increase bounded curve-tracking and progress rewards, with curvature-adaptive lookahead. 
\rule{\linewidth}{0.4pt}

\textbf{V2: Implementation of V1 LLM Feedback}

The following reward modifications were implemented based on the V1 evaluation:

\begin{itemize}
    \item Capped all penalty terms using Huber loss:
    \textit{max\_lane} = 2.0, \textit{max\_jerk} = 0.5,
    \textit{max\_oscillation} = 1.0.
    
    \item Increased the progress weight from 0.25 to 0.80.
    
    \item Reduced the smoothness weight from 0.80 to 0.25.
    
    \item Changed the oscillation penalty from count-based to a bounded
    rate-based measure in $[0,1]$.
    
    \item Introduced curvature-adaptive lookahead, using 6\,m for sharp
    curves and 12\,m for gentle curves.
    
    \item Added a curve-survival bonus for maintaining progress through
    curves without failure.
    
    \item Applied the V2 weights:
    \textit{lane} = 0.60, \textit{smoothness} = 0.25,
    \textit{curve} = 0.90, \textit{safety} = 1.00,
    \textit{progress} = 0.80.
\end{itemize}

\end{tcolorbox}
\newpage

\begin{tcolorbox}[
  colback=white,
  colframe=black,
  arc=2mm,
  title={Part of V2 Evaluation and V3 Reward Refinement}
]

\textbf{GPT-5.2 Evaluation:}

\textbf{Root Cause:}
\textit{The reward still does not create a strong, well-conditioned gradient toward the correct steady-state steering for curvature: curve tracking signals (lane/heading/curvature consistency) are either too weak, too saturated by caps, or not aligned with the control objective, so the agent can earn decent return via progress/survival while exhibiting oscillatory, reactive steering that eventually drifts off-road on sharper/longer bends.}

\textbf{V2 Reward Adjustments:}
\begin{itemize}
    \item \textit{lane\_keeping} = 0.75
    \item \textit{smoothness} = 0.45
    \item \textit{curve\_tracking} = 1.00
    \item \textit{safety} = 1.00
    \item \textit{progress} = 0.55
\end{itemize}

\textbf{Priority Fixes:}
\begin{enumerate}
    \item Make curve tracking ‘non-optional’
    
    \item Replace/augment oscillation shaping with action-derivative shaping that targets true instability
    
    \item Rebalance progress vs quality
\end{enumerate}

\textbf{summary:}
V1 fixed reward-shaping intentions but over-penalized (large negative returns) and reduced progress without improving outcomes. V2’s capping and rebalancing restored learnability and increased long runs and \texttt{good\_progress} frequency, but the policy still fails primarily from oscillatory, reactive curve steering and accumulated lateral drift. V3 should keep bounded penalties but strengthen curve-specific lane/heading/curvature-consistency shaping and gate progress/survival on curve-tracking quality to eliminate `progress despite bad tracking' solutions.
\rule{\linewidth}{0.4pt}
\textbf{V3: Implementation of V2 LLM Feedback}

The following reward modifications were implemented based on the V2 evaluation:

\begin{itemize}
    \item Added a curvature-consistency reward using a feedforward steering target based on waypoint angle.
    
    \item Added a counter-steer penalty for steering opposite to the curve direction when $|\texttt{wp\_angle}| > 10^\circ$.
    
    \item Gated progress using $\text{progress} \times \text{curve\_quality\_factor}$, where
    $\text{quality} = \exp(-k(\text{lane\_err} + \text{heading\_err}))$.
    
    \item Added a conditional survival bonus applied only when
    $\text{lane\_error} < 0.35$ and $\text{heading\_error} < 0.3$.
    
    \item Introduced a two-stage lane penalty consisting of Huber loss near the center and a boundary hinge for offset $> 0.5$.
    
    \item Added a curve-weighted heading-alignment reward.
    
    \item Increased the smoothness weight from 0.25 to 0.45.
    
    \item Increased the lane-keeping weight from 0.60 to 0.75.
    
    \item Increased the curve-tracking weight from 0.90 to 1.00, making it the highest priority.
    
    \item Reduced the progress weight from 0.80 to 0.55 while gating it on tracking quality.
    
    \item Applied the V3 weights:
    \textit{lane} = 0.75, \textit{smoothness} = 0.45,
    \textit{curve} = 1.00, \textit{safety} = 1.00,
    \textit{progress} = 0.55.
\end{itemize}
\end{tcolorbox}
\newpage

\newpage
\section{Additional Common-Sense Reasoning Examples}
\label{sec:common_sense_examples}

Additional examples of context-dependent rules generated by the common-sense
reasoner are provided to illustrate its behavior under combined driving
conditions.
\begin{figure}[H]
    \centering
    \begin{subfigure}{0.48\textwidth}
        \centering
        \includegraphics[width=\textwidth]{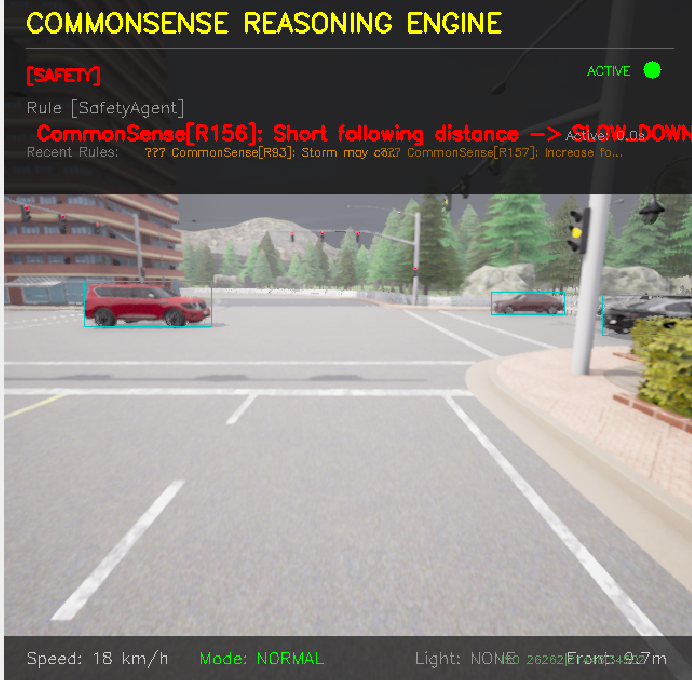}
        \caption{}
        \label{fig:following}
    \end{subfigure}
    \hfill
    \begin{subfigure}{0.48\textwidth}
        \centering
        \includegraphics[width=\textwidth]{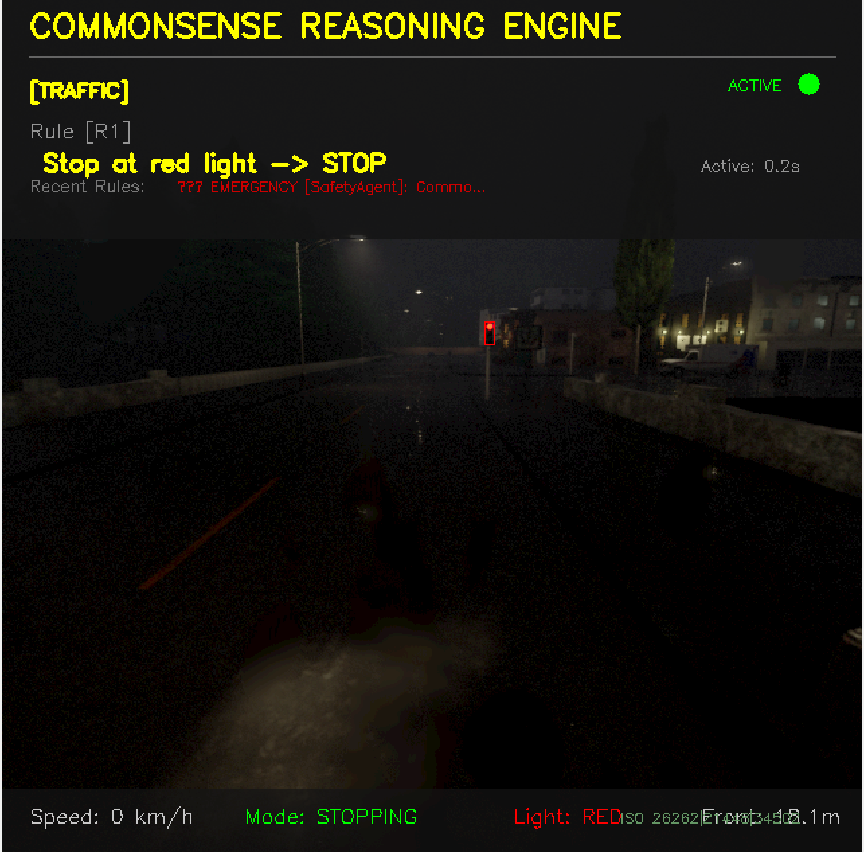}
        \caption{}
        \label{fig:traffic_light}
    \end{subfigure}
    \caption{Additional examples of common-sense reasoning in CARLA.
    (a) Short following distance with an approaching storm.
    (b) Red traffic light under heavy rain.}
    \label{fig:appendix_common_sense}
\end{figure}

\end{document}